\documentclass[fleqn,twoside]{article}
\usepackage{espcrc2}

\usepackage{graphicx}
\usepackage[figuresright]{rotating}

\newcommand{\AmS}{{\protect\the\textfont2
  A\kern-.1667em\lower.5ex\hbox{M}\kern-.125emS}}

\title{Forward Pion Production in Hadron Collisions at STAR}

\author{G.~Rakness\address[MCSD]{Penn State University/Brookhaven 
        National Laboratory, \\ 
        Physics 510-A, Upton, NY  11973},
        for the STAR Collaboration}
       
\begin{document}

\begin{abstract}
Measurements are reported of the production of high energy $\pi^0$ 
mesons at large pseudorapidity, coincident with charged hadrons at 
mid-rapidity, for proton+proton and deuteron+gold collisions at 
$\sqrt{s_{NN}}=200\ $GeV.  
The p+p cross section for inclusive $\pi^0$ production follows 
expectations from next-to-leading order perturbative QCD.  
A suppression of the back-to-back azimuthal correlations was 
observed in d+Au, qualitatively consistent with the gluon 
saturation picture.
Experimental uncertainties regarding the inclusive measurement are 
discussed.
\vspace{1pc}
\end{abstract}

\maketitle

\section{INTRODUCTION}

In contrast to the case for the nucleon, little is known
about the distribution of gluons in nuclei, especially in the
region of Bj\"{o}rken-$x<0.01$, where $x$ is the fraction of 
the nucleon's momentum carried by the parton.
Studies of deuteron(proton)+nucleus collisions at large center of mass 
energy ($\sqrt{s_{NN}}$=200 GeV) can provide constraints on the gluon 
density in heavy nuclei.
In the perturbative QCD explanation of particle
production at large pseudorapidity, a large-$x$ parton scatters from a 
low-$x$ parton and then fragments into the observed particle(s).
Forward charged particle production is found to be suppressed in
d+Au collisions \cite{brahms}, consistent with the expectation of
gluon saturation \cite{kharzeev-incl,jamal-2}, possibly indicating a
different mechanism for particle production.
Explanations of the suppression based on leading-twist pQCD 
calculations using a model of gluon shadowing have also been 
suggested \cite{vogt}. 
Further tests of the possible role played by gluon saturation at RHIC 
energies are provided by the study of particle correlations 
\cite{dis2004,monojet}.  
                                                                               
At $\sqrt{s}$=200 GeV and larger collision energies, there is 
quantitative agreement between NLO pQCD calculations and measured 
p+p cross sections at mid-rapidity \cite{phenix}.  
This agreement has been found to extend to $\pi^0$ production at 
$\langle \eta \rangle=3.8$ \cite{star-pi0}.  
Further tests of the underlying dynamics responsible for forward 
particle production can be obtained from the study of particle 
correlations.  
In particular, strong azimuthal correlations of hadron pairs are 
expected when particle production arises from $2 \rightarrow 2$ 
parton scattering.
                                                                               
This paper reports cross sections for forward inclusive $\pi^0$ 
production for p+p collisions at $\sqrt{s}$=200 GeV.
The azimuthal correlations between a forward $\pi^0$ 
($\langle \eta_{\pi} \rangle$=4.0) and mid-rapidity charged hadrons 
were studied.
In addition, exploratory studies with d+Au collisions at 
$\sqrt{s_{NN}}$=200 GeV are reported, and azimuthal correlations of 
hadron pairs are compared to those for p+p collisions.
The inclusive yields of $\pi^0$ mesons in p+p and d+Au collisions will
be forthcoming, and details pertinent to this analysis are presented 
here.
Forward $\pi^0$ production in d+Au collisions refers to observation 
of the $\pi^0$ in the direction of the incident deuteron.

\section{EXPERIMENT}

The Solenoidal Tracker At RHIC (STAR) is a multipurpose detector at
Brookhaven National Laboratory.  
One of its principal components is a time projection chamber 
used for tracking 
charged particles produced at $|\eta|<$1.2.
A forward $\pi^0$ detector (FPD) comprising $7\times 7$ matrices of 
$3.8\times 3.8\times 45$ cm$^3$ Pb-glass detectors (towers)
was installed $\approx 800\,$cm from the interaction region
near the beam pipe to detect high energy $\pi^0$ 
mesons with $3.3< \eta < 4.1$.  
Data were collected over two years of RHIC operations.
In the 2002 run, p+p collisions were studied with a prototype 
FPD \cite{star-pi0}.
In the 2003 run, p+p collisions were studied and exploratory 
measurements were performed with d+Au collisions.
\begin{figure}
  \centerline{
  \includegraphics[width=8.0cm,clip]{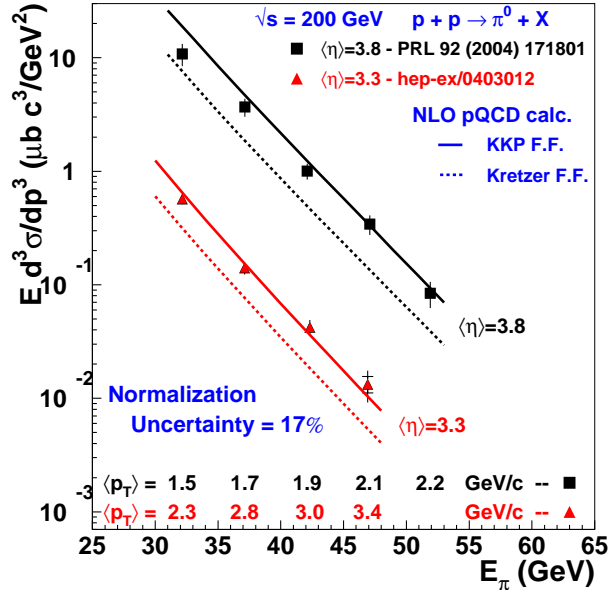}
   }   
  \caption{\label{fig:crosssec} 
    Inclusive $\pi^0$ production cross section versus 
    $\pi^0$ energy ($E_\pi$) at average pseudorapidities 
    ($\langle \eta \rangle$) 3.3 and 3.8.
    The inner error bars are statistical, and are smaller
    than the symbols for most points.  
    The outer errors combine these with $E_\pi$-dependent 
    systematic errors.  
    The curves are NLO pQCD calculations evaluated at $\eta=3.3$ 
    and 3.8 using different fragmentation functions.
}
\end{figure}

\section{DIFFERENTIAL CROSS SECTION}

The differential cross section for inclusive $\pi^0$ production for
$30 < E_\pi < 55\ $GeV at $\langle \eta \rangle=3.8$ was previously 
reported \cite{star-pi0}.
The event reconstruction and normalization methods were extended to 
allow measurement of the differential cross section at 
$\langle \eta \rangle=3.3$ \cite{eta3.3}.  
The results are shown in Fig. \ref{fig:crosssec} in comparison to NLO
pQCD calculations evaluated at $\eta$=3.3 and 3.8.
The NLO pQCD calculations are consistent with the data, in contrast to
$\pi^0$ data at lower $\sqrt{s}$ \cite{soffer}.  
The inclusive yield was also found to agree with predictions
of the PYTHIA event generator \cite{pythia}, which uses initial- and
final-state parton showers to model effects beyond leading order.
                                                                               
\begin{figure}
  \centerline{
  \includegraphics[width=8.0cm,clip]{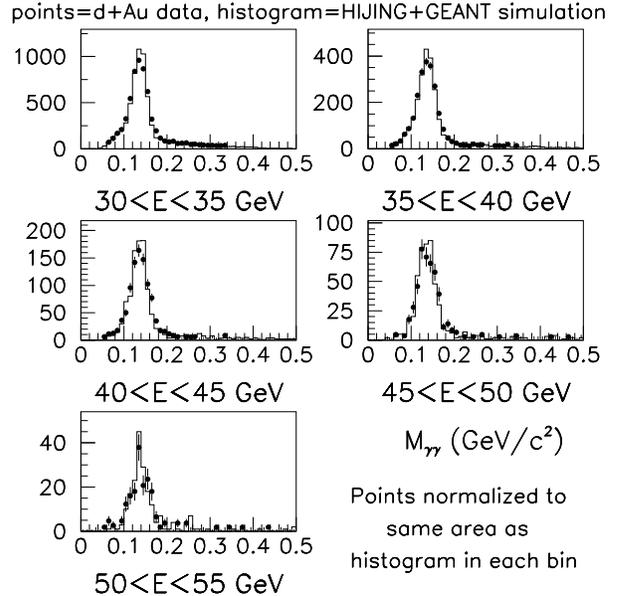}
   }   
  \caption{\label{fig:overlay} 
   Diphoton invariant mass spectra as a function of energy 
   deposited in the calorimeter for d+Au data and HIJING+GEANT 
   simulations.
   The histograms are reconstructed simulation events.
   The points are data with statistical errors, normalized to equal
   area in each bin.
}
\end{figure}
Nuclear effects on particle production are quantified by $R_{dAu}$,
the ratio of the inclusive yield of particles produced in d+Au 
collisions to p+p collisions.
The value of $R_{dAu}$ for forward $\pi^0$ mesons is expected to be 
smaller than was seen for negative hadrons \cite{brahms} due 
to isospin suppression of the $p+p\rightarrow h^-+X$ process \cite{gsv}.
We proceed to discuss considerations relevant to an accurate
determination of $R_{dAu}$ for forward $\pi^0$ production.

The data in Fig.~\ref{fig:crosssec} show that the cross section falls 
rapidly with $E_\pi$.  
Accurate yield determinations require accurate energy calibrations
and good energy resolution.
The energy calibration of the FPD is performed by 
reconstruction of the invariant mass of the $\pi^0$ meson, 
$M_{\gamma\gamma}$.
An analysis of the topology of the energy deposition, based on 
measured shower shapes \cite{showershape}, is used to reconstruct
the relative energy of the two photons and their separation 
($d_{\gamma\gamma}$).
The $\pi^0$ energy is taken to be the total energy in the 
calorimeter.
The gain of each tower is determined iteratively by associating 
$M_{\gamma\gamma}$ with the tower containing the most energy in the 
event.
After convergence, an energy-dependent correction is applied to 
account for material between the interaction 
region and the detector, which causes decay photons to shower prior to 
reaching the FPD.
It also corrects for the non-linearity implied by using 
digitizers with a finite number of bits.
The energy can be accurately associated with $M_{\gamma\gamma}$  
up to a point set by the systematics of reconstructing 
$d_{\gamma\gamma}$.
At 50\,GeV, $d_{\gamma\gamma}\approx 1.1\times$ the linear dimension 
of a tower.
The $M_{\gamma\gamma}$ distribution can be seen in Fig.~\ref{fig:overlay},
with resolution $\leq 20\,$MeV/c$^2$ from $30-55\,$GeV.
\begin{figure}
  \centerline{
  \includegraphics[width=8.0cm,clip]{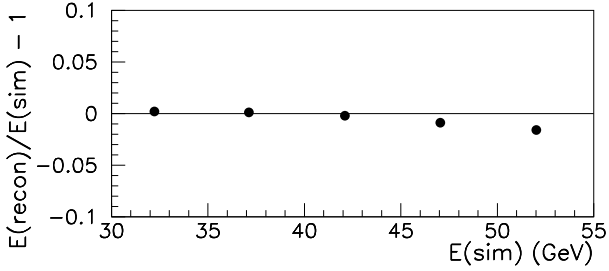}
   }   
  \caption{\label{fig:association} 
  Simulation study of the fractional difference of reconstructed 
  and simulated energy versus simulated $\pi^0$ energy.
  The energy calibration is determined to an accuracy
  of $\approx 1\%$.
  Systematics of reconstructing the di-photon opening angle are 
  seen beginning around $50\,$GeV.
}
\end{figure}

A simulation is used to model the events using the PYTHIA
(HIJING)\cite{hijing} generator for p+p (d+Au) collisions, together 
with a GEANT simulation to model the detector response, including 
intervening material.
After calibrating the simulation with the same technique as the data, 
the overlay of data and simulation for $M_{\gamma\gamma}$
can be seen in Fig.~\ref{fig:overlay}.
The simulation describes the data well for $M_{\gamma\gamma}$
as well as several other kinematic variables.
The simulations are used to determine the efficiency of $\pi^0$
detection.
As was observed in \cite{star-pi0}, the detection efficiency is
dominated by the geometrical acceptance of the calorimeter.

\begin{figure}
  \includegraphics[width=8.0cm,clip]{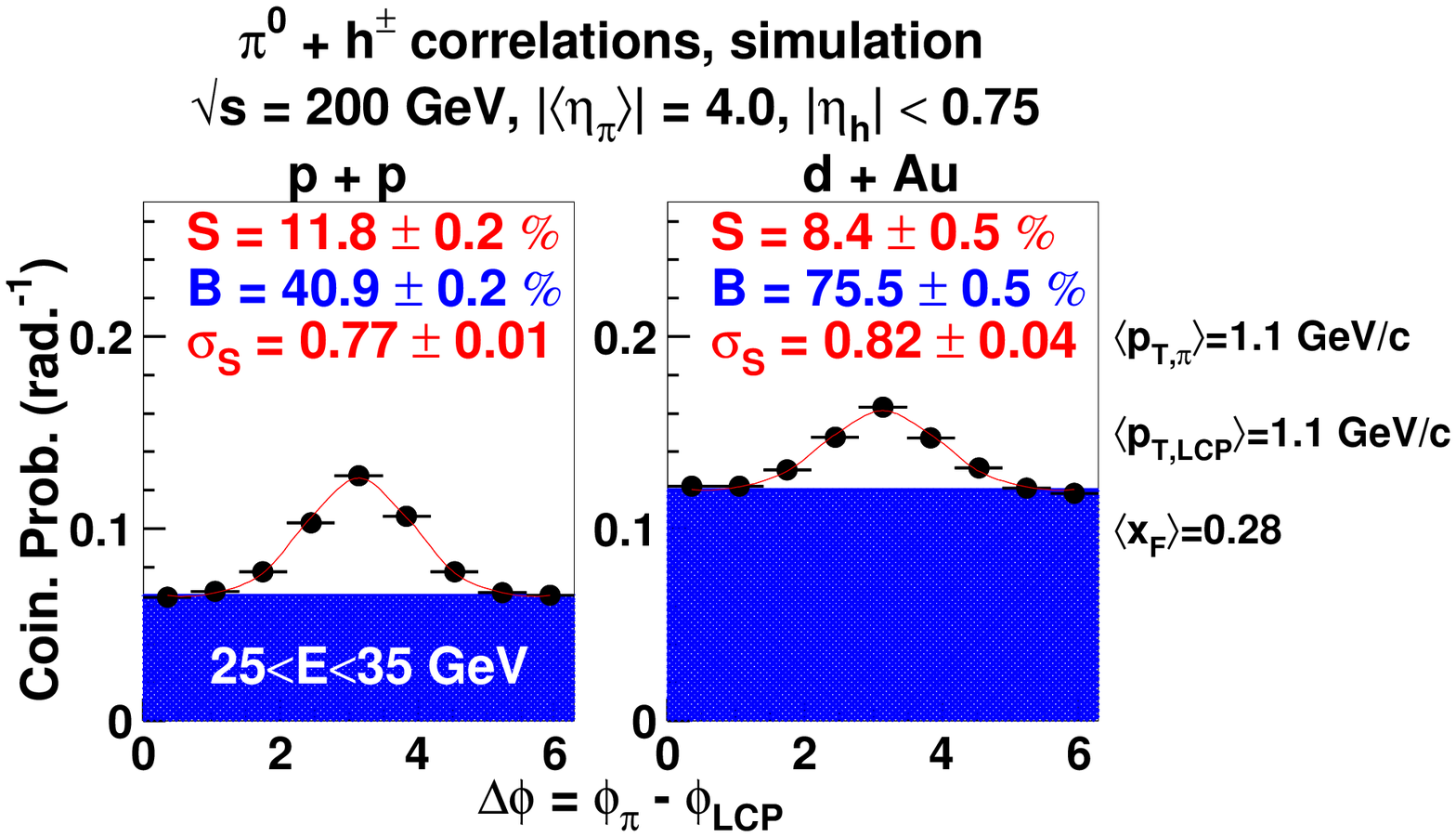}
  \includegraphics[width=8.0cm,clip]{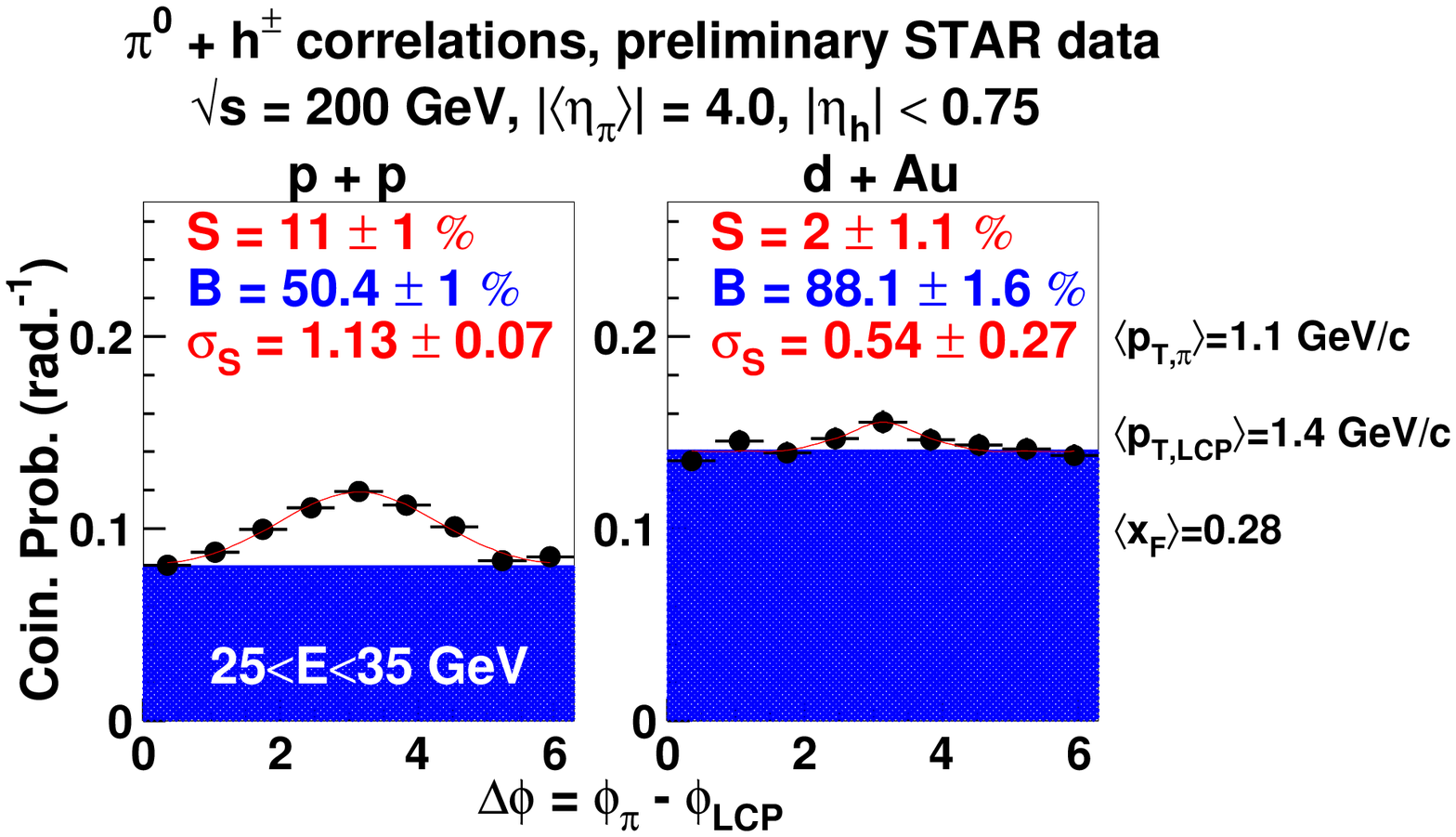}
  \caption{\label{fig:daucorr}
  Coincidence probability as function of azimuthal angle
  difference between a forward $\pi^0$ 
  and a midrapidity leading charged particle
  for p+p (left) and d+Au (right) collisions.
  The upper plots are simulation using PYTHIA 6.222 and HIJING 1.381
  described in the text, while the lower plots are data.}
\end{figure}
Using the simulation, a comparison of the kinematics of events that 
contain $\pi^0$ mesons with reconstructed variables gives an estimate 
of the accuracy of the analysis techniques used.
The fractional difference in energy between the generated and
reconstructed $\pi^0$ meson in bins of generated $\pi^0$ energy
can be seen in Fig.~\ref{fig:association}.
The uncertainty on the energy calibration is $\approx 1\%$, implying 
an uncertainty on the yield of $\approx 10\%$.

The data in Fig.~\ref{fig:crosssec} show that the cross 
section increases rapidly with $\eta$.  
Accurate yield determinations require accurate position determinations.
A Beam Position Monitor (BPM) is located in the vicinity of the FPD,
and has been accurately surveyed.
Relative to the BPM, the absolute position of the FPD is 
presently measured with an accuracy of 5\,mm, giving an absolute 
accuracy of $\delta\eta = 0.02$ at $\eta=4.00$.
This implies an uncertainty on the yield of 11\%.
                                                                               
\section{CORRELATIONS}

Correlations between a $\pi^0$ produced at large rapidity with large
Feynman $x$ and charged particles produced at 
midrapidity were studied with p+p collisions and d+Au 
collisions.
Details of the analysis, including detailed comparisons to 
simulations, may be found in Ref.~\cite{dis2004}.
The leading charged particle (LCP) analysis selects the mid-rapidity 
track with the highest $p_T>$0.5 GeV/c, and computes
the azimuthal angle difference $\Delta\phi=\phi_{\pi^0}-\phi_{LCP}$ 
for each event.
The normalized $\Delta\phi$ distributions are fit with the sum of a
constant and a Gaussian distribution centered at $\Delta\phi=\pi$.
The fit parameters are the area under the Gaussian 
(S), representing the azimuthally correlated coincidence probability; 
the uncorrelated coincidence probability (B); and the 
Gaussian width ($\sigma_S$).
The normalized $\Delta \phi$ distribution for d+Au collisions
is shown in comparison to 
p+p for both simulations and data in Fig.~\ref{fig:daucorr}. 
The simulations account for detector resolution and reconstruction
efficiency for both the forward $\pi^0$ and the midrapidity charged
particles.
                                                                             
We observe a large increase of B$_{\rm dAu}$ relative to B$_{\rm pp}$.  
The growth in B arises from additional nucleon-nucleon collisions
as the deuteron and Au beams interact.
For the data, $\sigma_S^{\rm dAu}$ is much smaller than 
$\sigma_S^{\rm pp}$, most likely reflecting the inadequacy of the 
functional form used to represent $\Delta\phi_{\rm dAu}$.  
The azimuthally correlated $\pi^0+h^\pm$ coincidence probability is 
smaller in d+Au collisions than in p+p collisions, qualitatively 
consistent with behavior predicted in a gluon saturation picture 
\cite{monojet}.  
Complete assessment of systematic errors is underway.               

\section{SUMMARY}

In summary, cross sections for the inclusive production of $\pi^0$
mesons in p+p collisions at $\sqrt{s}$=200 GeV, at $\langle \eta_{\pi}
\rangle$=3.3 and 3.8 are consistent with NLO pQCD calculations.
The azimuthal correlation between pairs of hadrons separated by large 
$\Delta\eta$ is described by PYTHIA \cite{pythia}, which uses leading 
order pQCD with parton showers.  
Agreement of these calculations with the inclusive cross section and 
di-hadron correlations suggest that forward $\pi^0$ production arises 
from partonic scattering at this collision energy.  
Exploratory studies of forward $\pi^0$ production in d+Au collisions 
suggest that the azimuthally correlated component of hadron pairs 
separated by large $\Delta\eta$ is suppressed relative to p+p 
collisions.  
More data for forward particle production and di-hadron correlations 
in d+Au collisions are required to reach a definitive conclusion 
about the possible existence of gluon saturation in the Au nucleus.


\begin{thebibliography}{9}
                                                                               
  \bibitem{brahms}
  I.~Arsene {\it et al.}, Phys. Rev. Lett. {\bf 93} 242303 (2004).
                                                                              
  \bibitem{kharzeev-incl} D.~Kharzeev, Y.\,V.~Kovchegov and
  K.~Tuchin, Phys.Rev. D {\bf 68} 094013 (2003).
                                                                              
  \bibitem{jamal-2} J.~Jalilian-Marian, {\mbox nucl-th/0402080}.
                                                                              
  \bibitem{vogt} R. Vogt, {\mbox hep-ph/0405060}.

  \bibitem{dis2004} A.~Ogawa for the STAR Collaboration,
         {\mbox nucl-ex/0408004}.
                                                                              
  \bibitem{monojet}
  D.~Kharzeev, E.~Levin and L.~McLerran,
  {\mbox hep-ph/0403271}.

  \bibitem{phenix} S.\,S.~Adler {\it et al.}, Phys. Rev. Lett. {\bf
  91}, 241803 (2003).

  \bibitem{star-pi0} J.~Adams, {\it et al.},
    Phys. Rev. Lett. {\bf 92} 171801 (2004).

  \bibitem{eta3.3} L.\,C.~Bland for the STAR Collaboration,
         {\mbox hep-ex/0403012}.

  \bibitem{soffer}  C.~Bourrely and J.~Soffer,
  Eur. Phys. J. C {\bf 36}, 371 (2004).
                                                                              
  \bibitem{pythia} T.~Sj\"{o}strand, {\it et al.},
  Comp. Phys. Commun. {\bf 135}, 238 (2001).

  \bibitem{gsv} V.~Guzey, M.~Strikman and W.~Vogelsang, 
         Phys. Lett. B{\bf 603}, 173 (2004).

  \bibitem{showershape} A.\,A. Lednev, Nucl. Instrum. Meth. A
      {\bf 366}, 292 (1995).

  \bibitem{hijing}
  X.~N.~Wang and M.~Gyulassy,
  Phys.\ Rev.\ D\ {\bf 44} 3501 (1991).

                                                                               
\end{thebibliography}
\end{document}